\documentclass[12pt]{article}
\usepackage{epsfig}
\usepackage{url}
\makeatletter
\def\alt{\mathrel{\mathpalette\gl@align<}}
\def\agt{\mathrel{\mathpalette\gl@align>}}
\def\gl@align#1#2{\lower.6ex\vbox{\baselineskip\z@skip\lineskip\z@
\ialign{$\m@th#1\hfil##\hfil$\crcr#2\crcr\sim\crcr}}}
\makeatother

\begin{document}
\begin{flushright}
MIFP-09-33 \\
UMD-PP-09-044 \\
IPPP/09/53, DCPT/09/106 \\
July, 2009
\end{flushright}
\vspace*{1.0cm}
\begin{center}
\baselineskip 20pt
{\Large\bf
Correlation between direct dark matter detection and Br($B_s \to \mu \mu$)
with a large phase of $B_s$-$\bar B_s$ mixing
} \vspace{1cm}

{\large
Bhaskar Dutta$^*$, Yukihiro Mimura$^\#$ and Yudi Santoso$^\dagger$}
\vspace{.5cm}

$^*${\it
Department of Physics, Texas A\& M University,
College Station, TX 77843-4242, USA
}\\
$^\#${\it
Department of Physics, University of Maryland,
College Park, MD 20742, USA
}\\
$^\dagger${\it
Institute for Particle Physics Phenomenology,\\
Department of Physics, University of Durham,
Durham, DH1 3LE, UK
}

\vspace{1.5cm}
{\bf Abstract}
\end{center}
We combine the analyses for flavor changing neutral current processes and dark
matter solutions in minimal-type supersymmetric grand unified theory (GUT) models,
SO(10) and SU(5), with a large $B_s$-$\bar B_s$ mixing phase and large $\tan
\beta$. For large $\tan \beta$, the double penguin diagram dominates the SUSY contribution
to the $B_s$-$\bar B_s$ mixing amplitude. Also, the Br($B_s \to\mu\mu$) constraint becomes
important as it grows as $\tan^6 \beta$, although it can still be suppressed by large pseudoscalar Higgs mass $m_A$. We investigate the correlation between $B_s \to\mu\mu$ and the dark matter direct
detection cross-section through their dependence on $m_A$. In the minimal-type of
SU(5) with type I seesaw, the large mixing in neutrino Dirac couplings results in
large lepton flavor violating decay process $\tau\to\mu\gamma$, which in turn sets upper bound on $m_A$. In the SO(10) case, the large mixing can be chosen to be in the Majorana
couplings instead, and the constraint from  Br($\tau\to\mu\gamma$) can be avoided.
The heavy Higgs funnel region turns out to be an interesting possibility in both cases and the direct dark matter
detection should be possible in the near future in these scenarios.

\thispagestyle{empty}

\bigskip
\newpage

\addtocounter{page}{-1}

\section{Introduction}
\baselineskip 18pt

Recently, CDF and D$\O$ collaborations
have announced the analysis
of the flavor-tagged $B_s \to J/\psi \phi$ decay. The decay width difference and the
mixing induced
CP violating phase, $\phi_s$, have been extracted from their analysis
\cite{Aaltonen:2007he}.
In the Standard Model (SM),
the CP violating phase is predicted to be
small,
$\phi_s = 2\beta_s \equiv 2\, {\rm arg}\, (-V_{ts}V_{tb}^*/V_{cs}V_{cb}^*) \simeq
0.04$.
However, the measured values of the phase are large:
\begin{eqnarray}
\phi_s ({\rm CDF}) &\in& [0.28,1.29]\ \ (68\%\, {\rm C.L.}), \\
\phi_s ({\rm D\O}) &=& 0.57^{+0.30}_{-0.24}({\rm stat})
                       {}^{+0.02}_{-0.07}({\rm syst}).
\end{eqnarray}
Combined data analyses
including the semileptonic asymmetry in the $B_s$ decay
indicate that the CP violating phase deviates about
$3 \sigma$ from the SM prediction \cite{Bona:2008jn}.
If this large phase still persists in the upcoming results from Fermilab,
it implies the existence of new physics (NP) beyond the SM
and that the new physics requires a flavor violation in $b$-$s$ transition
as well as a phase in the transition.

Supersymmetry (SUSY) is the most attractive candidate to build NP models.
As it is well known, SUSY models have a natural dark matter candidate which is the lightest
SUSY particle (LSP).
Besides, the gauge hierarchy problem can be solved and a natural aspect of
the theory  can be developed from the weak scale to the ultra high
energy scale.
In fact, the gauge coupling constants of the Standard Model gauge symmetries
can unify at a high scale using the  renormalization group equations
(RGEs) involving the particle contents of the minimal SUSY standard
model (MSSM). This indicates the existence of grand unified
theories (GUTs).
 The well motivated SUSY GUTs 
have always been subjects of intense experimental and theoretical investigations.
Identifying a GUT model, as currently is, will be a major focus of the upcoming experiments.

The nature of the flavor changing neutral currents (FCNCs)
and the CP violating phase is very important to test the existence of new physics 
beyond the standard model. 
In SUSY models, the SUSY breaking mass terms for squarks and sleptons
must be introduced, and they have sources of FCNCs and CP violation beyond the
Kobayashi-Maskawa
theory.
In general, the soft breaking terms generate too large FCNCs,
hence
flavor universality is often assumed
in squark and slepton mass matrices
to avoid large FCNCs  
in the meson mixings and lepton flavor violations (LFV) \cite{Gabbiani:1988rb}.
The flavor universality is expected to be realized by the Planck scale physics.
However,
even if universality is realized at a scale such as the GUT scale
or the Planck scale,
non-universality in the SUSY breaking sfermion masses is still generated
from the evolution of RGEs, 
and this can lead to a small flavor violating transitions,
which could be observed in the ongoing experiments.

In the MSSM 
with right-handed neutrinos,
the induced FCNCs from the RGE effects are not large in the quark sector,
while sizable effects can be generated in the lepton sector due to
the large neutrino mixing angles \cite{Borzumati:1986qx}.
Within GUTs, however, loop effects due to the large neutrino mixings
can induce sizable FCNCs also in the quark sector
since the GUT scale particles which connect quark and lepton sectors can propagate in the loops~\cite{Barbieri:1994pv}.
As a result, the patterns of the induced FCNCs
highly depend on the unification scenario
and the heavy particle contents.
Therefore, it is important to investigate
FCNC effects to obtain a footprint of the GUT models.
If quark-lepton unification is manifested
in a GUT model,
the flavor violation in $b$-$s$ transition
can be responsible for the large atmospheric neutrino mixing \cite{Moroi:2000tk},
and
thus, the amount of the flavor violation in $b$-$s$ transition
(the second and the third generation mixing),
which is related to the $B_s$-$\bar B_s$ mixing and its phase,
 has to be related to the $\tau \to \mu\gamma$ decay
\cite{Dutta:2006gq,Parry:2005fp,Dutta:2008xg,Hisano:2008df}
for a given particle spectrum.
The branching ratio of the $\tau \to \mu\gamma$ decay
is being measured at the $B$-factory,
and thus, the future results on LFV and from the ongoing measurement of the
phase of $B_s$-$\bar B_s$ mixing
will provide an important information to probe
the GUT scale physics.

In Refs.\cite{Dutta:2008xg,Dutta:2009iy},
two of us have studied the correlation between Br($\tau\to\mu\gamma$)
and $\phi_s$, the phase in $B_s$-$\bar B_s$ mixing,
comparing between SU(5) and SO(10) GUT models,
and investigated the observational constraints in these models
in order to decipher the GUT physics.
The flavor violation, originating from the
loop corrections via heavy particles,
can be characterized by the
CKM (Cabibbo-Kobayashi-Maskawa) quark mixing matrix and the
MNSP (Maki-Nakagawa-Sakata-Pontecorvo) neutrino mixing matrix,
as well as the size of the Yukawa couplings.
Since the CKM mixings are small, it is expected that the neutrino mixings
 dominate the source of FCNCs at low energy.
It is important to know whether
the large neutrino mixings originate from Dirac-type
or Majorana-type neutrino Yukawa couplings.
When the large neutrino mixings originate from the Dirac
neutrino Yukawa couplings in a particular GUT model,
the (squared) right-handed down-type squark mass matrix, $M_{\tilde D^c}^2$,
as well as the left-handed lepton doublet mass matrix, $M_{\tilde L}^2$,
 can
have  flavor non-universality.
When the large mixings originate from the Majorana Yukawa couplings,
the left-handed squark mass matrix, $M_{\tilde Q}^2$,
can also have  flavor non-universality in addition to the other sfermions.
In the minimal-type of SU(5) GUT,
the large neutrino mixings originate from Dirac neutrino couplings
provided that there is no fine-tuning in the seesaw neutrino matrix.
On the other hand, in the minimal-type of SO(10) GUT,
the large neutrino mixings can originate from Majorana-type couplings.
In general, since SU(5) is a subgroup of SO(10),
one can construct a model, a non-minimal-type of SU(5) GUT, where the neutrino mixings
originate from Majorana-type couplings.
Conversely, if we allow fine-tuning in the Yukawa coupling matrices,
Dirac neutrino Yukawa couplings can be  the source of
the large mixings even in the SO(10) model.
Actually, there is a little ambiguity to determine the minimal SU(5) or SO(10) GUT
model
because the very minimal versions of the GUT models have problems with phenomenology
and a slight modification is needed.
(That is why we call our models ``minimal-type''.)
Here, we call a typical boundary condition as minimal-type SU(5) GUT condition
when the off-diagonal elements of $M_{\tilde D^c}^2$ and $M_{\tilde L}^2$
are correlated due to the Dirac neutrino couplings in the model.
Another type of  boundary condition where the $M_{\tilde Q}^2$ is  correlated to
$M_{\tilde D^c, \tilde U^c}^2$ and $M_{\tilde L}^2$ due
to the Majorana coupling in SO(10) model is called the minimal-type SO(10) GUT
boundary condition.
The large phase of $B_s$-$\bar B_s$ mixing in combination with the other flavor violating
processes,
can tell us which type of boundary condition is preferable.

We analyzed the case of a large $\tan\beta$
(which is the ratio of the vacuum expectation values of up- and down-type Higgs fields)
in the Ref.\cite{Dutta:2009iy}.
In large $\tan\beta$ case,
the so-called double penguin contribution \cite{Hamzaoui:1998nu,Buras:2001mb}
can dominate the SUSY contribution to the $B_s$-$\bar B_s$ mixing amplitude
over the box contribution
unless the pseudo Higgs field is heavy.
When the double penguin contribution is enhanced by a smaller pseudo Higgs field mass,
the $B_s \to \mu\mu$ decay \cite{Choudhury:1998ze,Buras:2001mb}
is also enhanced close to its experimental bound \cite{Foster:2004vp}.
In other words, if the large phase of $B_s$-$\bar B_s$ mixing
originates from the double penguin contribution,
the $B_s \to \mu\mu$ decay should be observed very soon,
and therefore it is worthwhile to examine the constraints to see if
the large phase is really generated from the double penguin contribution.
An important constraint to obtain a large phase of $B_s$-$\bar B_s$ in GUT models
comes from the experimental bound of $\tau\to\mu\gamma$ decay.
Due to the quark-lepton unification,
when the flavor violation of $b$-$s$ transition is large,
the $\tau$-$\mu$ flavor violation is expected to be large as well.
This is significant especially in the minimal-type of SU(5) model.
As a result,
a wide region of the parameter space is excluded.
This result is important to distinguish among the solutions of
the dark matter relic density as measured by Wilkinson Microwave Anisotropy
Probe (WMAP) \cite{Komatsu:2008hk} in the context of SUSY dark matter. Since imposing supersymmetric solution to the dark matter content of the universe puts a tremendous constraint on the SUSY model parameter space, it is interesting to see whether the constraints from the flavor violating processes can be satisfied by the dark matter allowed regions. The allowed parameter space, satisfying  both of these constraints, can then be probed directly at the large hadron collider (LHC).
In this paper, we assume the lightest neutralino as dark matter in the GUT models and combine the dark matter analyses with the flavor constraint analyses.
Consequently, some of the solutions for the neutralino dark matter are disfavored,
and the
so-called funnel solution (in which the neutralinos annihilate through the heavy Higgs bosons)
turns out to be an interesting one for the large phase of $B_s$-$\bar B_s$ mixing and  large $\tan\beta$ case considered here.
In the solution which satisfies all the constraints, the branching ratio of the $B_s \to \mu\mu$ decay  is predicted,
and  we will show that it is in the range to be observed soon.
In scenarios with large $\tan\beta$, the dark matter direct detection
can be correlated to the branching ratio of the $B_s\to\mu\mu$ decay since both are
enhanced by a small pseudoscalar Higgs mass \cite{Ellis:2006jy}.
We investigate this correlation for the case with large phase of $B_s$-$\bar B_s$
mixing.

The paper is organized as follows:
In section 2, we describe the FCNC sources in SUSY GUT models.
The two typical boundary conditions in both SU(5) and SO(10) models
are considered.
In section 3, we describe the SUSY contributions of $B_s$-$\bar B_s$ mixing amplitudes,
and the constraints from the other FCNC modes.
We also discuss the solutions of the WMAP dark matter relic density which
can be allowed in this scheme.
In section 4, we show our numerical results on  both SU(5) and SO(10) GUT models.
Section 5 is devoted to conclusion and remarks.

\section{FCNC sources in SUSY GUTs}

In SUSY theories, the SUSY breaking terms can be sources of
 flavor violations.
In general, it is easy to include  flavor violating terms  by hand
 since the SUSY
breaking masses with flavor indices are parameters in the MSSM.
However, large FCNCs
are induced if these parameters are completely general~\cite{Gabbiani:1988rb}.
Therefore, as a minimal assumption of the SUSY breaking, 
universality of the scalar masses is often considered.
This means that all the SUSY breaking (squared) scalar masses are equal
to $m_0^2$ and all the scalar trilinear couplings are proportional
to the Yukawa couplings (with the coefficients are universal to be $A_0$) at a
unification scale.

Even if the universality is assumed, 
non-universality in the scalar masses at the weak scale is generated by the evolution of
the theory from the GUT scale down to the lower scale via the RGEs.
As we have mentioned in the introduction, in the MSSM with right-handed neutrinos ($N^c$) the induced FCNCs from the RGE effects are not large in the quark sector while sizable effects can be generated in the lepton sector due to the large neutrino mixings \cite{Borzumati:1986qx}.
The sources of FCNCs in this model are the Dirac neutrino couplings.
In GUT models, the left-handed lepton doublet ($L$) and the right-handed down-type
squarks ($D^c$)
are unified in $\bar{\bf 5}$, and
the Dirac neutrino couplings can be written as $Y_\nu \bar{\bf 5} N^c H_{\bf 5}$.
As a result, non-universality in the SUSY breaking mass matrix for $D^c$
is generated from the colored-Higgs and right-handed neutrino loop diagrams,
and flavor violation in the quark sector can then also be generated from the Dirac
neutrino couplings \cite{Barbieri:1994pv,Moroi:2000tk}.

The light neutrino mass matrix can be written as
\begin{equation}
{\cal M}_\nu^{\rm light} = f \langle \Delta_L \rangle - Y_\nu M_R^{-1} Y_\nu^{\rm T}
\langle H_u^0 \rangle^2,
\end{equation}
where $\Delta_L$ is an SU(2)$_L$ triplet, and $f$ is a Majorana coupling $\frac12
LL\Delta_L$.
The second term is called type I seesaw term \cite{Minkowski:1977sc}.
If the type I seesaw term dominates the light neutrino mass,
the Dirac neutrino coupling must have large mixings to explain the large neutrino
mixings
in the basis where the charge-lepton Yukawa coupling matrix $Y_e$ is diagonal.
On the other hand, when the first term (triplet term) dominates (type II seesaw
\cite{Schechter:1980gr}),
the Majorana coupling must have large mixings.
Distinguishing these two cases is very important in order to understand the source
of FCNCs in the GUT models.

The triplet contribution in the type II seesaw is
natural in the framework of SO(10) GUT models \cite{Babu:1992ia}.
In the SO(10) models,
all matter species are unified in the spinor representation ${\bf 16}$.
Since the right-handed neutrino is also unified to other matter fields, the
 neutrino Dirac Yukawa coupling does not
have large mixings 
in the minimal-type of SO(10) models,
and the proper neutrino masses with large mixings
can be generated from the 
Majorana couplings $\frac12 f LL \Delta_L$.
The $f$ coupling is unified to the ${\bf 16}\,{\bf 16}\,\overline{\bf 126}$
term
which also includes Dirac Yukawa couplings for fermions,
and thus the model is predictive \cite{Dutta:2004wv}.
If any of the decomposed fields from $\overline{\bf 126}$ is lighter than
the unification scale,
the flavor non-universality for squarks and sleptons is generated.
It is then possible that the loop corrections generate
the flavor violation for the left-handed quark doublet $(Q)$,
the right-handed up-type quark $(U^c)$ and the right-handed charged-lepton $(E^c)$,
in addition to $D^c$ and $L$.

We parameterize the non-universality in the squark and slepton mass matrices due to
the loop corrections
as
\begin{equation}
M_{\tilde {F}}^2 = m_{0}^2 [{\bf 1} - \kappa_F U_F {\rm diag} (k_1,k_2,1)
U_F^{\dagger}],
\end{equation}
where $F = Q, U^c, D^c, L, E^c$.
The quantity $\kappa_F$ denotes the amount of the off-diagonal elements
and it depends on the sfermion species.
The unitary matrices $U_F$
is equal to the neutrino mixing matrix in a limit.
We note that the unitary matrices $U_F$ should be defined in the basis
where charged-lepton Yukawa matrix, and down-type quark Yukawa matrix are diagonal
in order to calculate the flavor violating processes such as $\tau \to \mu\gamma$,
and $B_s$-$\bar B_s$ mixing.
In the minimal SU(5) GUT where only $H_{\bf 5}$ and $\bar H_{\bar{\bf 5}}$
couple to the fermions by renormalizable terms,
$U_{D^c}$ is exactly same as $U_L$ and has large mixing angles,
while $U_{Q,U^c,E^c}$ have small mixings relating the CKM mixings.
In general, fermion mass matrices come from the sum of the Yukawa terms, and 
the equality of $U_{D^c}$ and $U_L$ can be completely broken
when there are cancellations among the minimal Yukawa term and
additional Yukawa terms. 
Here, we consider a model
where the (nearly) equality between $U_{D^c}$ and $U_L$ (especially for 
23 mixing angle of them) is maintained
as a ``minimal-type" assumption.
The assumption is natural if there is a dominant Yukawa contribution
and corrections to fit realistic masses and mixings are small.
In the minimal-type of SO(10) model, all $U_F$ can have large mixings
responsible for the neutrino mixings.
The detail physical interpretation of this parameterization is given in
\cite{Dutta:2009iy,Dutta:2006zt}.
When the Dirac neutrino Yukawa coupling $Y_\nu$ or the
Majorana coupling $f$ is hierarchical,
we obtain $k_1,k_2 \ll 1$ and
then the 23 element of the sfermion mass matrix
is $-1/2 m_0^2 \kappa \sin 2\theta_{23} e^{i \alpha}$.
The magnitude of the FCNC between 2nd and 3rd generations
is controlled by $\kappa \sin2\theta_{23}$,
where $\theta_{23}$ is the mixing angle in the unitary matrix.
The phase parameter $\alpha$ also originates from the unitary matrix,
and it will be the origin of a phase of the FCNC contribution.

It is interesting that the flavor violation pattern in the lepton sector
and the quark sector can depend on the SO(10) symmetry breaking vacua.
Actually, in order to forbid a rapid proton decay,
the quark flavor violation should be larger than the lepton flavor violation
among the symmetry breaking vacua \cite{Dutta:2007ai}.
Namely, it is expected that
$\kappa_{Q}$, $\kappa_{U^c}$, and $\kappa_{D^c}$ are much larger than
$\kappa_L$ and $\kappa_{E^c}$.
For example, if only the Higgs fields $({\bf 8},{\bf 2}, \pm1/2)$
are light compared to the breaking scale (which is the most suitable case),
one obtains $\kappa_Q = \kappa_{U^c} = \kappa_{D^c}$,
and only quark flavor violation is generated, while the lepton flavor violation is not generated.
On the other hand, when the flavor violation is generated
from the minimal-type of SU(5) vacua with type I seesaw,
the quantities $\kappa$'s have relations as $\kappa_{L} \sim \kappa_{D^c}$, and
$\kappa_Q, \kappa_{U^c},\kappa_{E^c} \sim 0$, effectively.
Actually, when we take the threshold effect into account,
it is expected that $\kappa_{L}$ is always larger than $\kappa_{D^c}$
since the right-handed Majorana mass scale is less than the scale of colored Higgs
mass.
Therefore, the existence of $b$-$s$ transition indicated by the experimental results
in Fermilab predicts the sizable lepton flavor violation in the minimal-type of
SU(5) model.
Thus, if the results of large $B_s$-$\bar B_s$ phase is really an evidence of NP,
the GUT models are restricted severely \cite{Parry:2005fp,Dutta:2008xg,Hisano:2008df}.
Therefore, investigating the quark and lepton flavor violation is very important to
decipher
the GUT symmetry breaking especially when the $B_s$-$\bar B_s$ phase is
large~\cite{Dutta:2008xg}.

\section{$B_s$-$\bar B_s$ mixing and direct dark matter detection}

Let us briefly see the phase of $B_s$-$\bar B_s$ mixing.
We use the model-independent parameterization of the NP contribution:
\begin{equation}
C_{B_s} e^{2i\phi_{B_s}} = M_{12}^{\rm full}/M_{12}^{\rm SM},
\end{equation}
where `full' means the SM plus NP contribution,
$M_{12}^{\rm full} = M_{12}^{\rm SM}+ M_{12}^{\rm NP}$.
The NP contribution can be parameterized by two real parameters $C_{B_s}$
and $\phi_{B_s}$.
The time dependent CP asymmetry ($S = \sin \phi_s$) in $B_s \to J/\psi \phi$
is dictated by the argument of $M_{12}^{\rm full}$ :
 $\phi_s = - {\rm arg} M_{12}^{\rm full}$,
and thus 
$\phi_s = 2(\beta_s - \phi_{B_s})$.
It is important to note  that large SUSY contribution is still allowed
even though  the mass difference of  $B_s$-$\bar B_s$ \cite{Abulencia:2006ze}
is kept fairly consistent with the SM prediction.
This is
because the mass difference, $\Delta M_{B_s}$, can be just  twice the absolute value
of  $M_{12}^{\rm full}$.
The consistency of the mass difference between the SM prediction
and the experimental measurement just means $C_{B_s} \sim 1$,
and a large $\phi_{B_s}$ is still allowed.
For example, when $C_{B_s} \simeq 1$,
the phase $\phi_{B_s}$ is related as
$2\sin \phi_{B_s} \simeq A_{s}^{\rm NP}/A_s^{\rm SM}$,
where $A_s^{\rm NP,SM} = | M_{12}^{\rm NP,SM} |$.
The argument of $M_{12}^{\rm NP}$, being free
in GUT models, is due to the phase in off-diagonal elements in
SUSY breaking mass matrix (in the basis where $Y_d$ is a real diagonal matrix),
and one can choose an appropriate value for the new phase in the NP contribution.
Therefore, the experimental data constrains $A_s^{\rm NP}/A_s^{\rm SM}$,
and therefore, $\kappa \sin2\theta_{23}$ is constrained for a given SUSY particle
spectrum
when the phase of $B_s$-$\bar B_s$ is large.

In the MSSM with flavor universality, the chargino box diagram dominates
the SUSY contribution to $M_{12}(B_s)$.
In the general parameter space of the soft SUSY breaking terms,
the gluino box diagram can dominate the SUSY contribution for a lower $\tan\beta$
(i.e. $\tan\beta \alt 30$).
The gluino box contribution is enhanced if both left- and right-handed down-type squark
mass matrices have off-diagonal elements \cite{Dutta:2006gq},
and
therefore,
it is expected that the SUSY contribution to the $B_s$-$\bar B_s$ mixing amplitude
is large for the SO(10) model with type II seesaw,
compared to the minimal-type of SU(5) model \cite{Dutta:2008xg}.

The box diagram does not depend on $\tan \beta$ (ratio of the vacuum expectation values
of two Higgs fields) explicitly,
whereas, the flavor changing Higgs interaction (through so-called Higgs penguin
diagram)
directly depend on $\tan\beta$,
and the double Higgs penguin contribution to the $B_s$-$\bar B_s$ mixing
can become more important
than the box diagram when $\tan\beta$ is large
and there is an off-diagonal element in the right-handed
down-type squark mass matrix \cite{Hamzaoui:1998nu,Buras:2001mb}.
We note that
the off-diagonal elements in the left-handed squark mass matrix is less important
in order to generate a sizable double penguin contribution.
This is because the chargino loop can generate the left-handed Higgs penguin
contribution.
Therefore, even in the minimal-type of SU(5) model,
the double penguin contribution can be sizable when $\tan\beta$ is large.
When the off-diagonal elements of left-handed squark mass matrix are generated,
the left-handed flavor changing contribution to different processes (e.g., $b\to
s\gamma$)
can be modified.

The $B_s \to \mu\mu$ decay can be generated by a single Higgs penguin
diagram \cite{Choudhury:1998ze,Buras:2001mb}.
The decay amplitude is proportional to the muon Yukawa coupling, and
thus the amplitude is proportional to $\tan^3 \beta$.
Therefore, the branching ratio is proportional to $\tan^6\beta$.
Since it can be generated by a single left-handed penguin diagram, this decay occurs
even in the universal SUSY breaking models like the mSUGRA (minimal supergravity)
\cite{Dedes:2001fv}.
The current bound of the branching ratio is Br($B_s \to \mu\mu$) $< 4.7 \times
10^{-8}$~\cite{Aaltonen:2007kv}.
When $\tan\beta$ is large, this bound gives an important constraint to the parameter
space
\cite{Parry:2005fp,Foster:2004vp,Ellis:2006jy,Ellis:2005sc}.
In other words, one would  expect that the $B_s \to \mu\mu$ decay will be observed
very soon.

When the lepton flavor violation is correlated to the flavor violation
in the right-handed down-type squark as in the minimal-type of SU(5) model,
the $\tau\to \mu\gamma$ decay will give us the most important constraint
to obtain the large $B_s$-$\bar B_s$ phase \cite{Dutta:2008xg,Hisano:2008df}.
Furthermore, the squark masses are raised  much more compared to
the slepton masses due to the gaugino loop contribution
since the gluino is heavier compared to the Bino and the Wino at low energy,
and thus the lepton flavor violation will be more sizable compared to the quark
flavor violation.
The current experimental bound of the branching ratio of $\tau\to\mu\gamma$
is \cite{Hayasaka:2007vc}
\begin{equation}
{\rm Br}(\tau\to\mu\gamma) < 4.5 \times 10^{-8}.
\end{equation}
In order to allow for a large phase in the $B_s$-$\bar B_s$ mixing in the
minimal-type of SU(5) model,
a large flavor-universal scalar mass (often called $m_0$) at the cutoff scale
is preferable.
The reasons are as follows.
The gaugino loop effects are flavor invisible
and they enhance the diagonal elements of the scalar mass matrices
while keeping the off-diagonal elements unchanged.
If the flavor universal
scalar masses at the cutoff scale become larger,
both Br($\tau\to\mu\gamma)$ and $\phi_{B_s}$ are suppressed.
However, Br($\tau\to\mu\gamma)$ is much more suppressed compared to $\phi_{B_s}$
for a given $\kappa \sin2\theta_{23}$
because the low energy slepton masses are sensitive to $m_0$
while the squark masses are not so sensitive
due to the gluino loop contribution to their masses.

When $\tan\beta$ is large, the $\tau\to\mu\gamma$ constraint
is relaxed for a large $B_s$-$\bar B_s$ phase,
because the double-penguin contribution to the $B_s$-$\bar B_s$ mixing is
proportional to $\tan^4\beta$ while the $\tau\to\mu\gamma$ is proportional
to $\tan^2\beta$.
However, the $B_s\to \mu\mu$ constraint becomes important in this case
since it is proportional to $\tan^6\beta$.
As a result,
the branching ratio of $B_s \to \mu\mu$
decay will have a lower bound in a given large $\tan\beta$ and SUSY spectrum
when the phase of $B_s$-$\bar B_s$ mixing is large.
This is because as follows:
The double penguin contribution to the $B_s$-$\bar B_s$ mixing
and the amplitude of $B_s \to \mu\mu$ are inversely proportional to $m_A^2$,
where  $m_A$ is a CP odd Higgs mass.
For a given $\tan\beta$ and a large phase of $B_s$-$\bar B_s$
mixing,
a larger $\kappa$ value is needed if $m_A$ is supposed to be larger.
Then, the parameter space is excluded by $\tau\to\mu\gamma$ constraint,
due to the approximate relation Br($\tau\to\mu\gamma) \propto \kappa^2$.
Therefore, $m_A$ has an upper bound, and $B_s\to \mu\mu$
have a lower bound.

Such constraints in the minimal-type of SU(5) have an impact on the neutralino
dark matter which satisfies the recent WMAP result
of relic density.
The dark matter constraint is mostly satisfied in the minimal supergravity (mSUGRA) models by the following three scenarios.
\begin{enumerate}
        \item The stau-neutralino coannihilation region.
        \item The (nearly) Higgsino region (i.e. lightest neutralino has a large Higgsino
component).
        \item The funnel region (i.e. the neutralinos annihilate through heavy Higgs bosons
pole).
\end{enumerate}
At first, the stau-neutralino coannihilation region is not favored
because the lighter stau is relatively light (almost degenerate with neutralino) in the region.
When the stau is light, the $\tau\to\mu\gamma$ constraint will be severe
and  a large phase of $B_s$-$\bar B_s$ mixing is excluded.
Secondly, the small Higgsino mass is not very favored either
because it also enhances the $\tau\to\mu\gamma$ amplitude
unless the sleptons are very heavy.
Such heavy sleptons do not explain the muon $g-2$ anomaly \cite{g-2}.
The third solution (funnel region)
is interesting in the current scheme to have a large phase of $B_s$-$\bar B_s$ mixing.
In the funnel region, the lightest neutralino mass $M_{\tilde \chi^0_1}$
is twice the mass of the heavy Higgs bosons ($2 M_{\tilde \chi^0_1} \simeq m_A$).
In the models that we consider, the masses of heavier CP even Higgs boson and
the CP odd Higgs boson are nearly degenerate.
As we have mentioned, $m_A$ has an upper bound for a given
parameter space,
and as result, the mass of the lightest neutralino
is bounded from above.
In the funnel region,
$m_A$ is almost fixed for a given gaugino mass.
In the double penguin contribution of $B_s$-$\bar B_s$ amplitude
and the $B_s \to \mu\mu$, $m_A$ is a dominant parameter,
and thus the branching ratio of $B_s \to \mu\mu$ can be predicted.

The Higgs mass $m_A$ is also important for
the spin-independent scattering cross-section
since it gets the dominant contributions from the Higgs exchange diagrams.
In much of the parameter space, the t-channel
Higgs exchanges ($h$, $H$) dominate the proton-neutralino cross-section
$\sigma_{\tilde{\chi}_{1}^{0}-p}$.
The spin-independent scattering cross-section  can be written as~\cite{kamionkowski}:
\begin{equation}
\sigma_{\tilde\chi^0_1-p}\simeq \frac{4}{\pi} m_p^4|
(A^uf_u/m_u+A^cf_c/m_c+A^tf_t/m_t)+(A^df_d/m_d+A^sf_s/m_s+A^bf_b/m_b)|^2,
\end{equation}
where, $f_q \equiv \langle p| m_q \bar q q | p\rangle/m_p$,
and
$f_u \simeq 0.027$; $f_d \simeq 0.039$; $f_s \simeq 0.36$; $f_c=f_b=f_t \simeq 0.043$~\cite{Ellis:2005mb}.
The down- and up-type quark Higgs amplitudes are
\begin{eqnarray}
A^{d,s,b} = \frac{g_2^2 m_{d,s,b}}{4 M_W} \left( - \frac{\sin \alpha}{\cos \beta}
\frac{F_h}{m_h^2} + \frac{\cos \alpha}{\cos \beta} \frac{F_H}{m_H^2} \right),
\label{quark-Higgs-amplitude}
 \\
A^{u,c,t} = \frac{g_2^2 m_{u,c,t}}{4 M_W} \left( \frac{\cos \alpha}{\sin \beta}
\frac{F_h}{m_h^2} + \frac{\sin \alpha}{\sin \beta} \frac{F_H}{m_H^2} \right),
\end{eqnarray}
where $g_2$ is the $SU(2)$ gauge coupling, the Higgs mixing angle $\alpha$ is usually small $\sim 0.1$ and
\begin{eqnarray}
F_h = (N_{12}-N_{11} \tan \theta_W)(N_{14} \cos{\alpha} + N_{13} \sin{\alpha}),
 \\
F_H = (N_{12}-N_{11} \tan \theta_W)(N_{14} \sin{\alpha} - N_{13} \cos{\alpha}).
\end{eqnarray}
Here $N_{1i}$ are the mixing amplitudes for the lightest neutralino 
$\tilde\chi^0_1$ among the
Bino ($\tilde B$), Wino ($\tilde W$) and the two Higgsinos ($\tilde H_1,$ $\tilde
H_2$):
\begin{equation}
\tilde\chi^0_1=N_{11}\tilde B+N_{12}\tilde W+N_{13}\tilde H_1+
N_{14}\tilde H_2.
\end{equation}
We see that the down-type quark Higgs amplitude, Eq.(\ref{quark-Higgs-amplitude}),
has larger contribution to the
cross-section rather than the up-type ones $A^{u,c,t}$ due to larger $f_s$.
It is also apparent that the cross-section increases for smaller value of the heavy Higgs
mass, $m_H$, which scales with $m_A$ and smaller values of $\mu$ which increases $N_{13}$.

Therefore, $m_A$ is also important for the proton-neutralino elastic scattering
cross-section,
and this is particularly the case for models with non-universal Higgs masses
\cite{Accomando:1999eg,Ellis:2003eg,Baer:2008ih}.
Thus, there is a correlation
between the direct detection of the Milky Way dark matter
and Br($B_s \to \mu\mu$) \cite{Ellis:2006jy}. Furthermore, the dark matter cross-sections are also affected by its particle contents, i.e., whether the lightest neutralino is gaugino or Higgsino
dominated.

The current highest sensitivity of direct detection is about $5 \times 10^{-8}$ pb for neutralino mass $\alt 100$~GeV
\cite{Angle:2007uj,Ahmed:2008eu}. This is expected to
increase to $2 \times 10^{-9}$ pb 
soon \cite{Aprile:2009yh} for neutralino mass $\alt 100$~GeV (for the neutralino mass we
have considered in this paper the expected limit should be around $5 \times 10^{-9}$~pb). For the models we
consider, this constraint can exclude the parameter space with small $m_A$ and/or
small $\mu$.

\section{Numerical results}

In order to illustrate the features described
in the previous section,
we plot the figures
when the NP/SM ratio of
the $B_s$-$\bar B_s$ amplitude is 0.5, $A_s^{\rm NP}/A_s^{\rm SM} = 0.5$,
and the absolute value of the full amplitude is same as SM amplitude, $C_{B_s} = 1$.
Under these choices, one can obtain that $|2\phi_{B_s}|$ is about 0.5 (rad).
We consider that the SUSY breaking Higgs squared masses,
$m_{H_u}^2$ and $m_{H_d}^2$, are not related to other scalar masses
in order to make $m_A$ and $\mu$ free parameters,
since these two parameters are important for Higgs penguin contribution
and the proton-neutralino cross-section.

\begin{figure}[htbp]
 \center
 \includegraphics[viewport = 0 10 570 570,width=8cm]{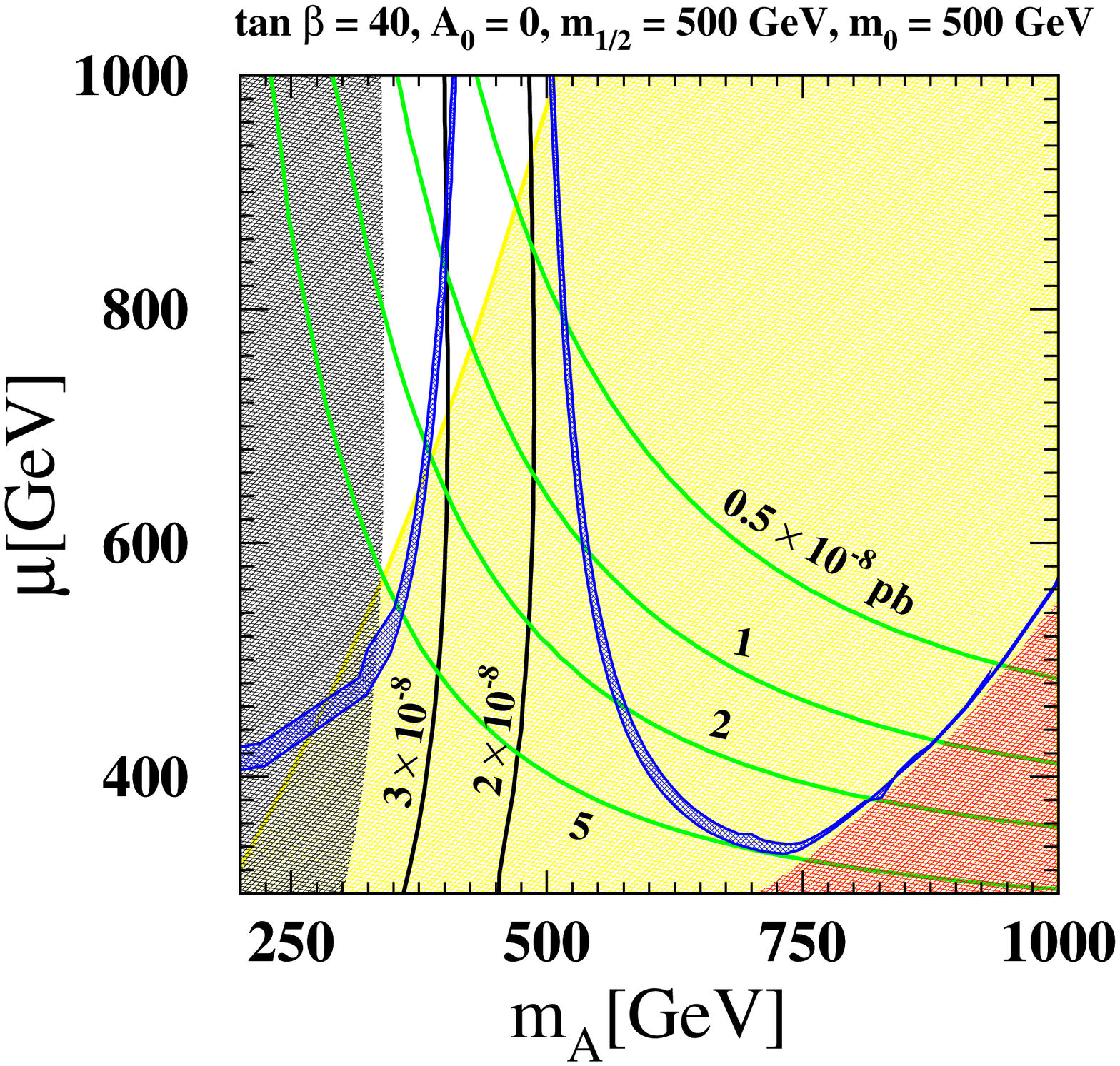}
 \includegraphics[viewport = 0 10 570 570,width=8cm]{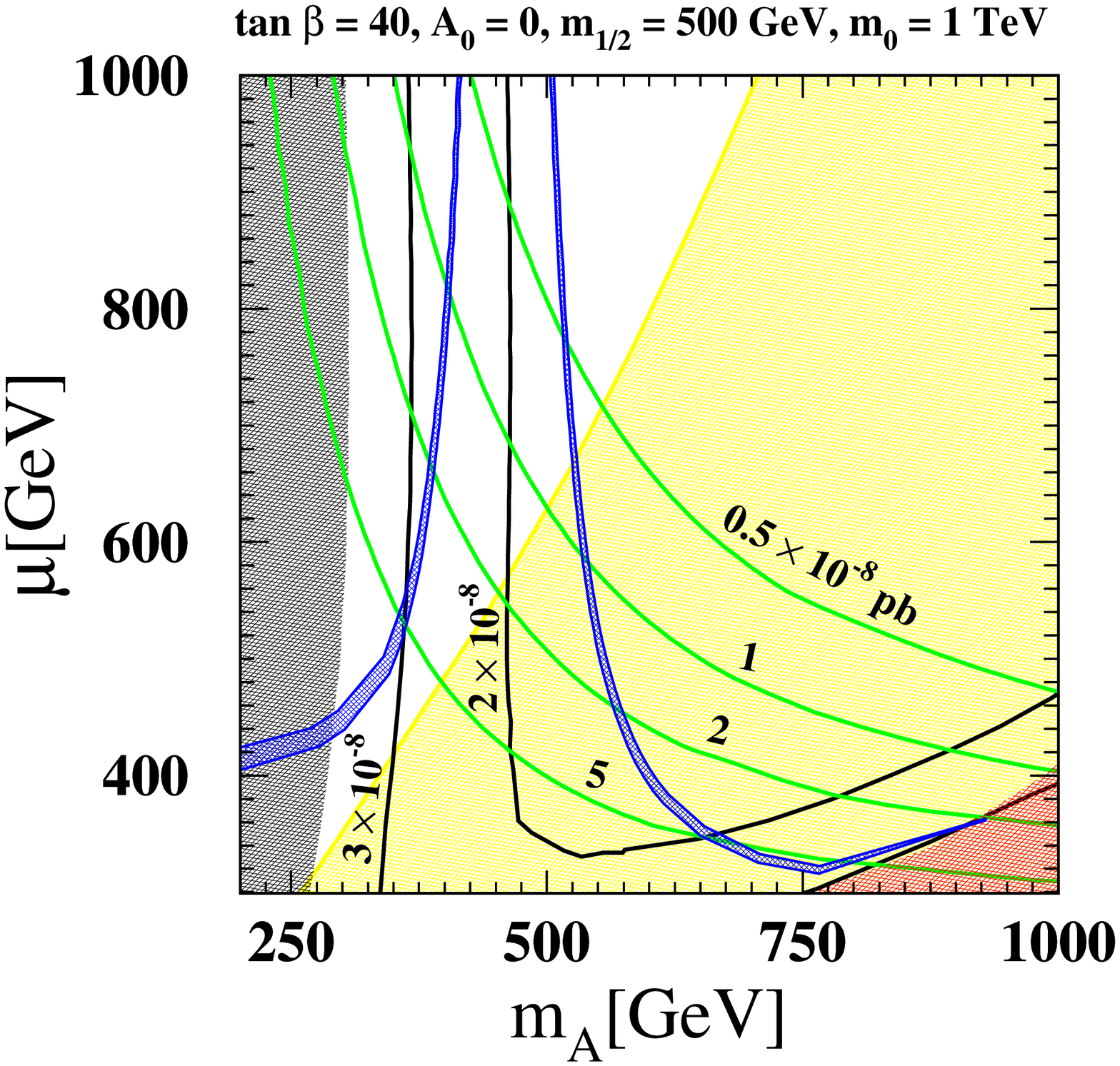}
 \includegraphics[viewport = 0 10 570 570,width=8cm]{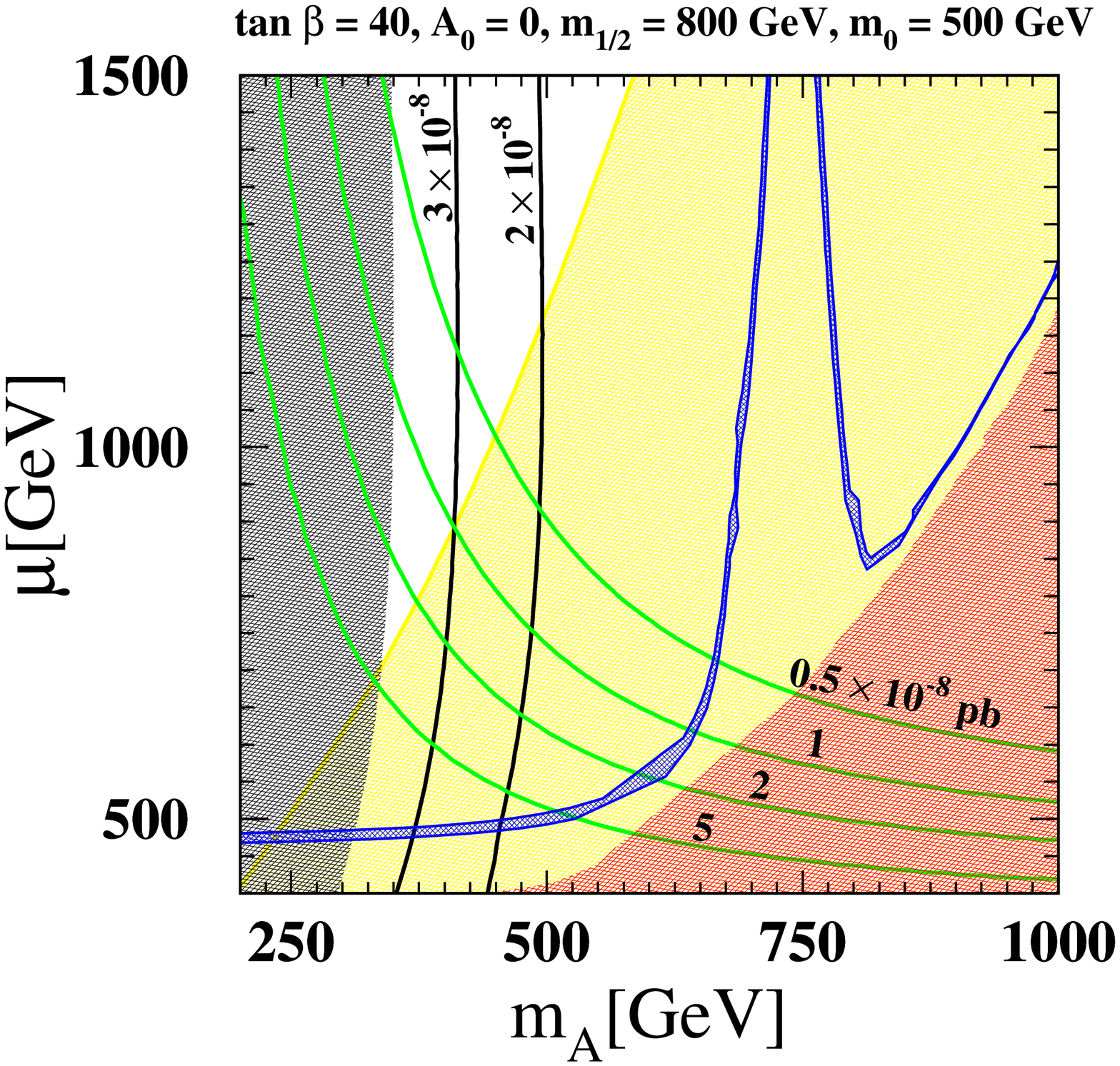}
 \includegraphics[viewport = 0 10 570 570,width=8cm]{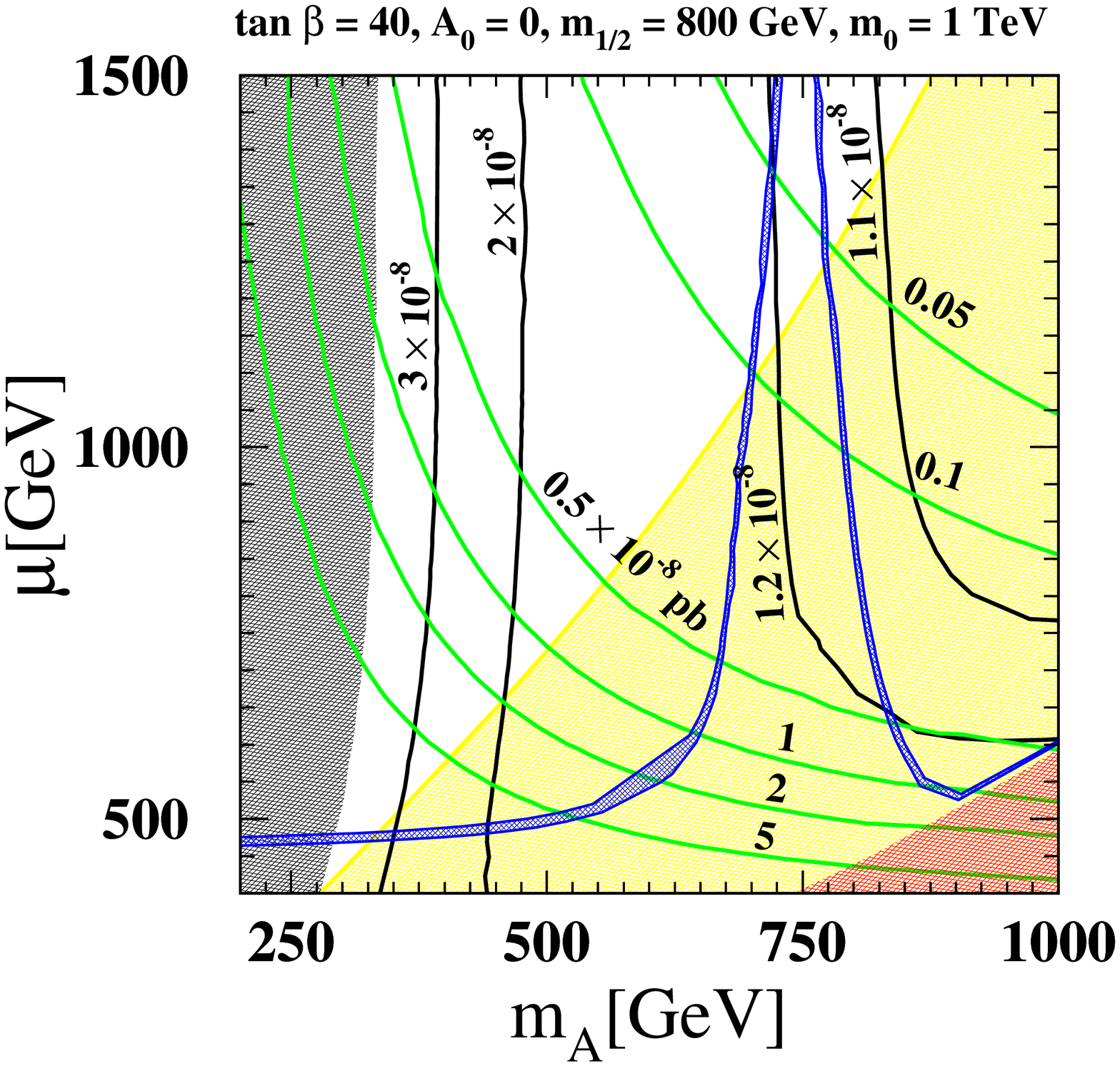}
 \caption{
The $m_A - \mu$ plane for minimal-type of SU(5) with non-universal Higgs masses, for
$\tan \beta = 40$, $A_0 = 0$ and $(m_{1/2}, m_0) =$ (a) (500,500)~GeV, (b)
(500,1000)~GeV, (c) (800,500)~GeV and (d) (800,1000)~GeV respectively.
}
\end{figure}

Figure 1 is drawn in the case of minimal-type of SU(5) model.
We choose the $\kappa$ values for the non-universality
to be same for left-handed sleptons and right-handed down-type squarks,
for simplicity.
To draw, we choose
$\tan\beta =40$,
and the universal trilinear scalar coupling at GUT scale is zero ($A_0=0$).
The unified gaugino mass $m_{1/2}$ is chosen to be 500 GeV and 800 GeV,
and the sfermion masses at GUT scale
is chosen to be 500 GeV and 1 TeV. These ranges of  mass parameters can be probed at the LHC.
We plot Br($B_s \to \mu\mu$) and the proton-neutralino spin independent cross-section ($\sigma_{\chi-p}$) contours using black and green lines respectively.
The blue strips are the 2-std region of the WMAP dark matter relic density, assuming
that the entire dark matter content is made of neutralino.
The gray shaded region is excluded by the experimental bound of Br($B_s\to\mu\mu$).
The red shaded region (bottom-right corner)
corresponds to the stau LSP and hence is disallowed by the dark matter requirement.
The yellow shaded region is excluded by the
experimental bound of Br($\tau\to\mu\gamma$).
As one can see from the figures,
the $\tau\to\mu\gamma$ bound is relaxed
for a larger sfermion mass $m_0$.
For a larger gaugino mass $m_{1/2}$,
the WMAP allowed strips
for the funnel region (i.e., the vertical strips for $2M_{\tilde \chi_1^0}\sim m_A$) shift to
the right.
As mentioned in the previous section,
the stau-neutralino coannihilation region
(close to the red shaded region)
and the small Higgsino mass (i.e. small $\mu$)
is excluded by the bound of Br($\tau\to\mu\gamma$).
On the other hand, the funnel region
is still allowed.
When $m_{1/2}=800$ GeV,
the funnel regions (for $\mu< 1$ TeV) shift to the region excluded by
$\tau\to\mu\gamma$.
In order to allow these regions, $\mu > 2$ TeV and $\mu > 1.2$ TeV are needed
for $m_0$= 500 GeV and 800 GeV respectively.
As a consequence,
if $\mu$ is restricted to be less than 1 TeV,
$m_{1/2}$ is bounded and
then the cross-section for the direct dark matter detection
is bounded from below.
Note that the neutralino mass, which depends mostly on $m_{1/2}$, in these plots, is large (about 200~GeV for (a) and (b), and 320~GeV for (c) and (d)). At these
masses, the current sensitivity of the direct detection experiments (as can be seen
in Fig.~4 of \cite{Angle:2007uj} and
Fig.~4 of \cite{Ahmed:2008eu}) is still lower
than what is needed to exclude more of the parameter space. With the expected
increase of the sensitivity by an order of magnitude in the near future, this
constraint would become more severe.

\begin{figure}[tbp]
 \center
 \includegraphics[viewport = 0 10 570 570,width=8cm]{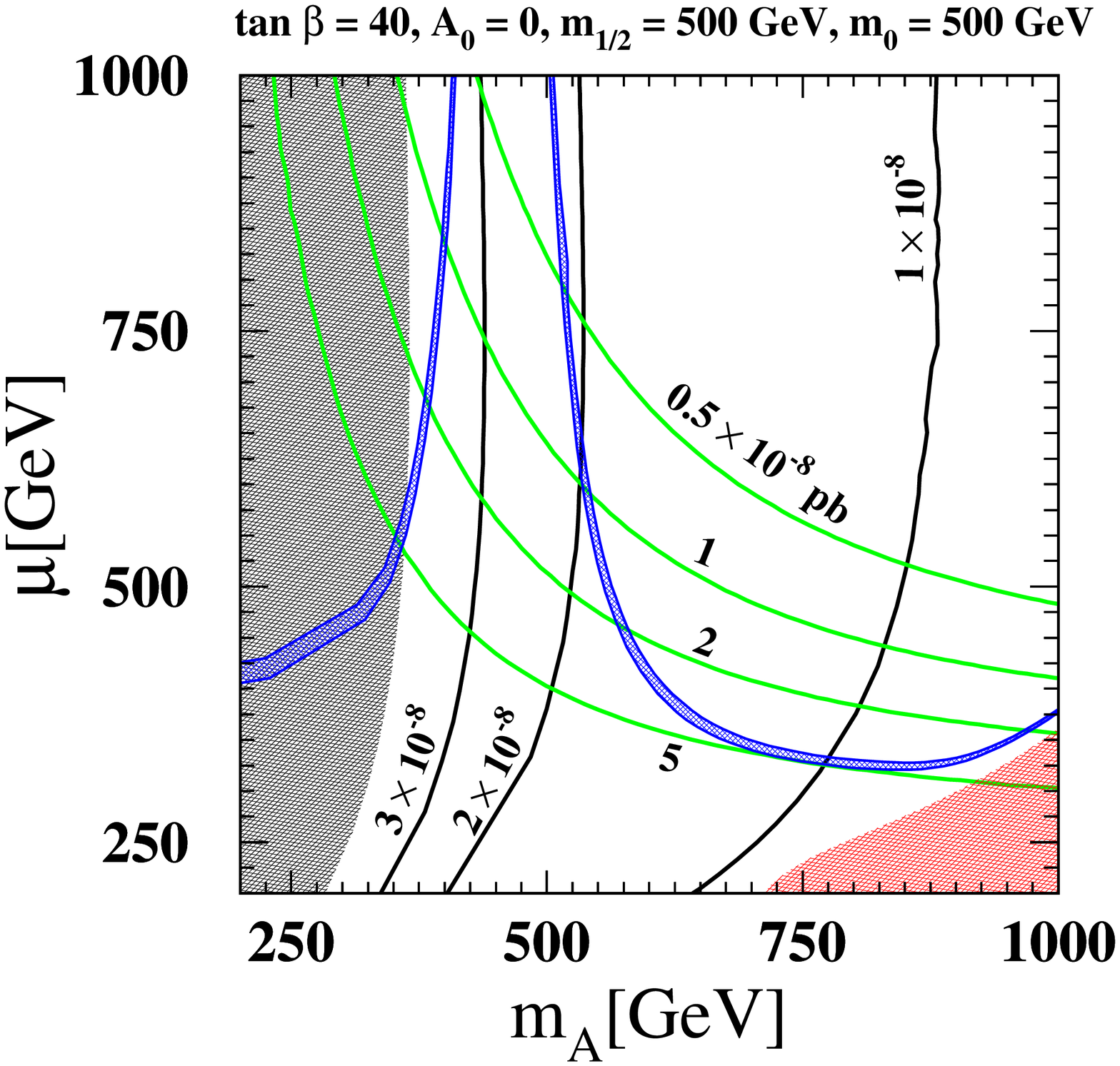}
 \includegraphics[viewport = 0 10 570 570,width=8cm]{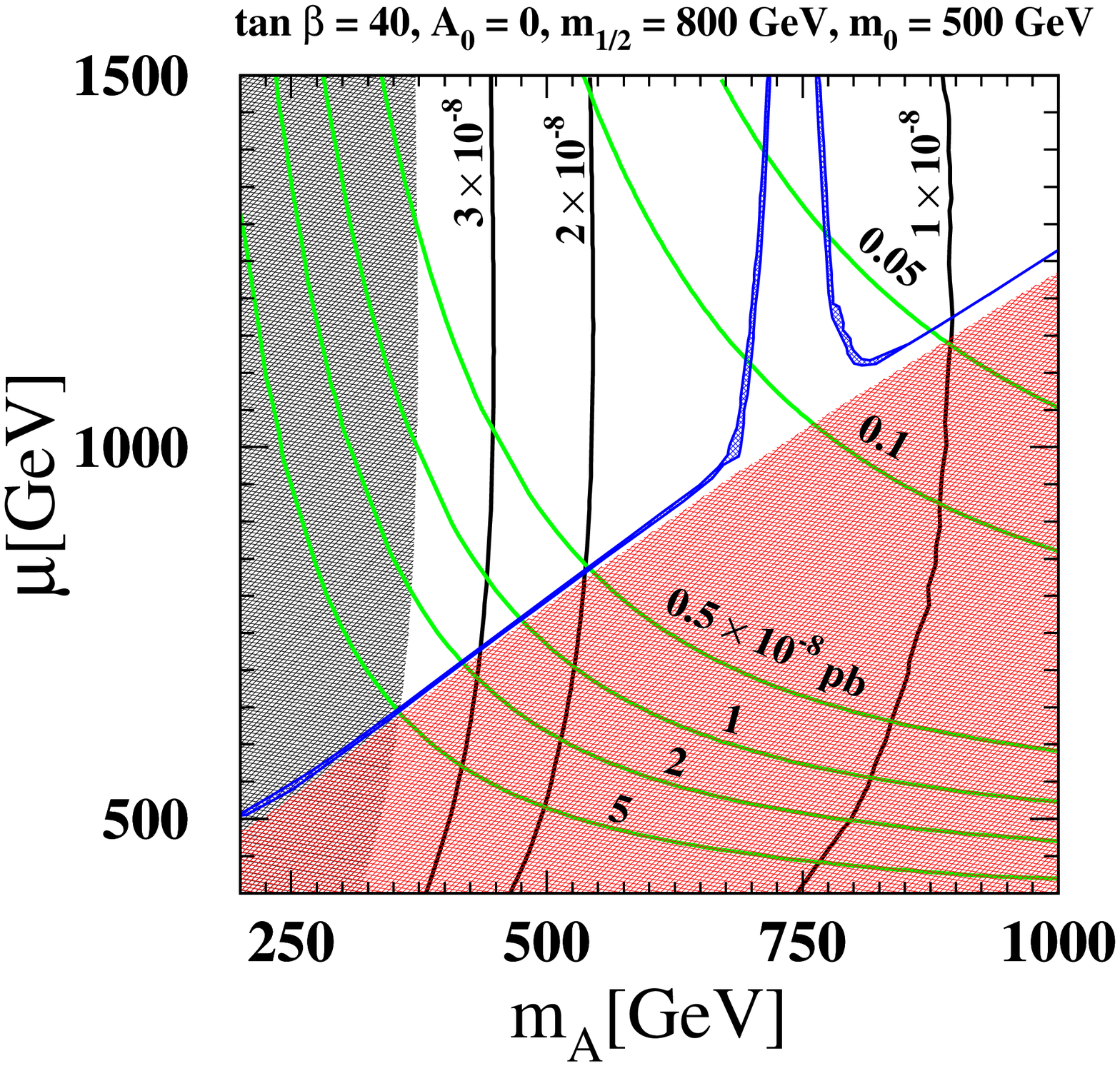}
 \caption{
The $m_A - \mu$ plane for minimal type SO(10) models with non-universal Higgs
masses, for $\tan \beta = 40$, $A_0 = 0$ and $(m_{1/2}, m_0) =$ (a) (500,500)~GeV,
and (b) (800,500)~GeV  respectively.
}
\end{figure}

Figure 2 is drawn for the case of SO(10) boundary conditions in the same way as for
figure 1.
We choose the kappa values to be same
for both left- and right-handed squarks.
As is mentioned, depending on the SO(10) breaking vacua,
the flavor non-universality in the slepton mass matrices
can be reduced,
and
we choose them to be zero to escape from the $\tau\to\mu\gamma$ bound.
With this choice, then, there is no upper bound for $m_A$.
Therefore,
if Br($B_s \to\mu\mu$) turns out to be small ($\alt 10^{-8}$) and $\tan\beta$ large,
we should adopt the SO(10) model.
As we have mentioned, in the SU(5) case, when we take  both the dark matter relic density and $\tau\to\mu\gamma$ bound into account,
$\mu$ needs to be larger for a larger $m_{1/2}$,  
and consequently the direct detection cross-section would be too small to be probed at the ongoing XENON~100 experiment, as can be seen from figures 1(c,d). 
In the SO(10) case, on the other hand,
the direct detection cross-section even for larger values of $m_{1/2}$ (e.g. figure 2(b))
can still be large and could be probed very soon.
This is, in part, because the stau coannihilation and the  Higgsino dark matter solutions are also allowed along with the funnel region.
If we find evidence for these solutions from the LHC, then the SO(10) model would be preferred.

We comment that the boundary condition for the left-handed squark mass
and right-handed up-type squark masses are different from the
previous SU(5) boundary condition.
Therefore, the RGE running for the SUSY breaking Higgs masses are different,
and thus the stau LSP regions (red/pink shaded regions) for the SO(10) figures are different from the SU(5) ones.

In Figure 3,
we show the correlation between
the Br($B_s\to \mu\mu$) and the proton-neutralino cross-section
for SU(5) and SO(10) boundary conditions.
The data points are picked up from the figure 1 and 2
for $m_0= m_{1/2}=500$ GeV,
and are sliced for fixed values of the Higgsino mass $\mu$.
The small circles represents the solution for the WMAP relic density.
Since $m_A$ is almost determined for the funnel region of the
WMAP solution
and the $\mu$ dependence of Br($B_s\to\mu\mu$)
for a given phase of $B_s$-$\bar B_s$ mixing is
not large,
Br($B_s\to\mu\mu$) is predictable as one can see from the figures.
The proton-neutralino cross-section, on the other hands, depends on the Higgsino
mass hence varies with $\mu$. The current experimental sensitivity is still  low,
but if one of the solutions is correct, the direct detection of the dark matter
should be possible in the near future.
Note, however, that if the neutralino contributes only a part of the dark matter content (i.e.,
regions between two circles where the neutralino relic density is lower than the
WMAP), the neutralino direct detection rate would be scaled down. In this case, 
we will need even higher sensitivity for the direct detection to exclude the parameter space regions.

\begin{figure}[tbp]
 \center
 \includegraphics[viewport = 0 10 570 400,width=8cm]{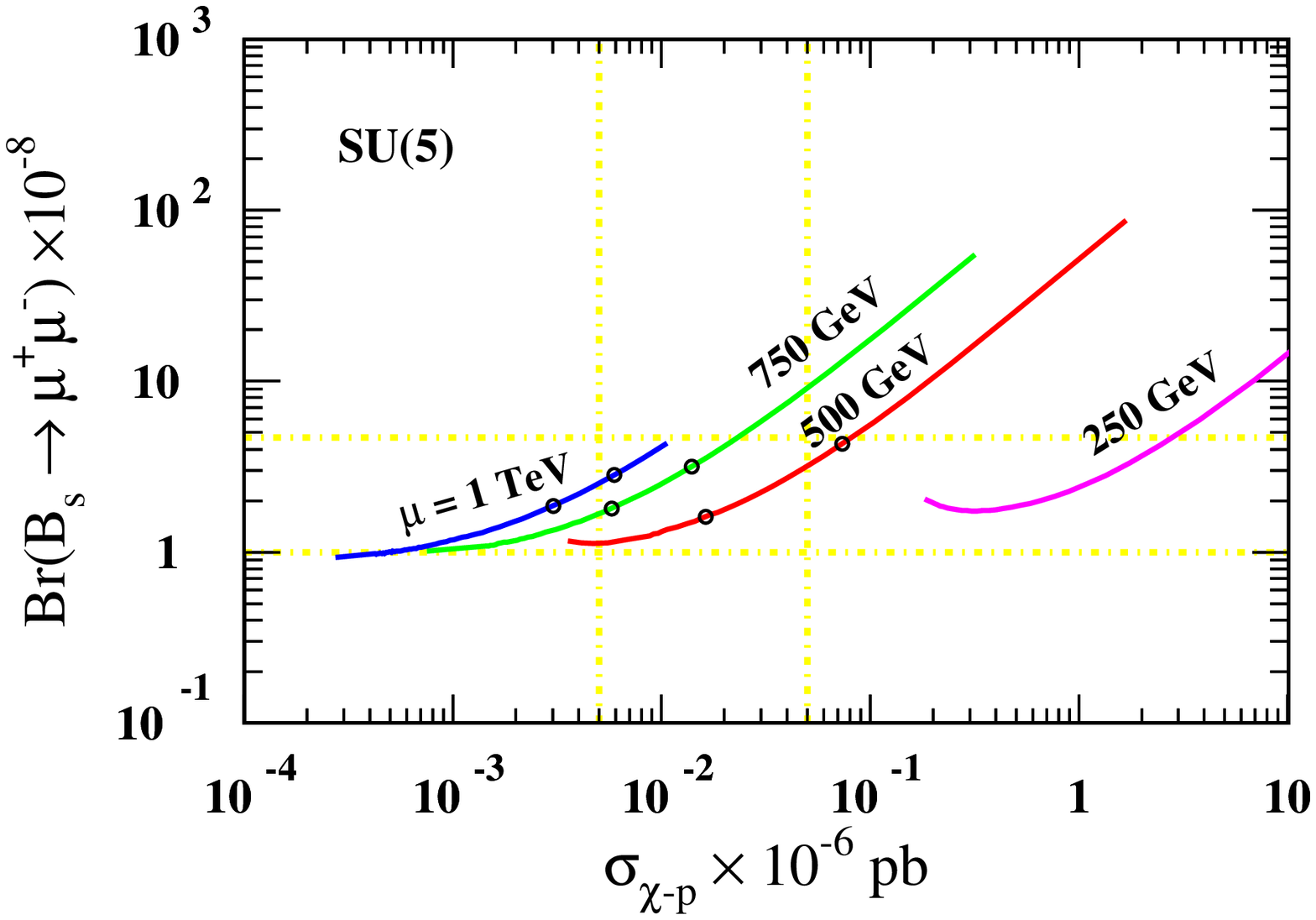}
 \includegraphics[viewport = 0 10 570 400,width=8cm]{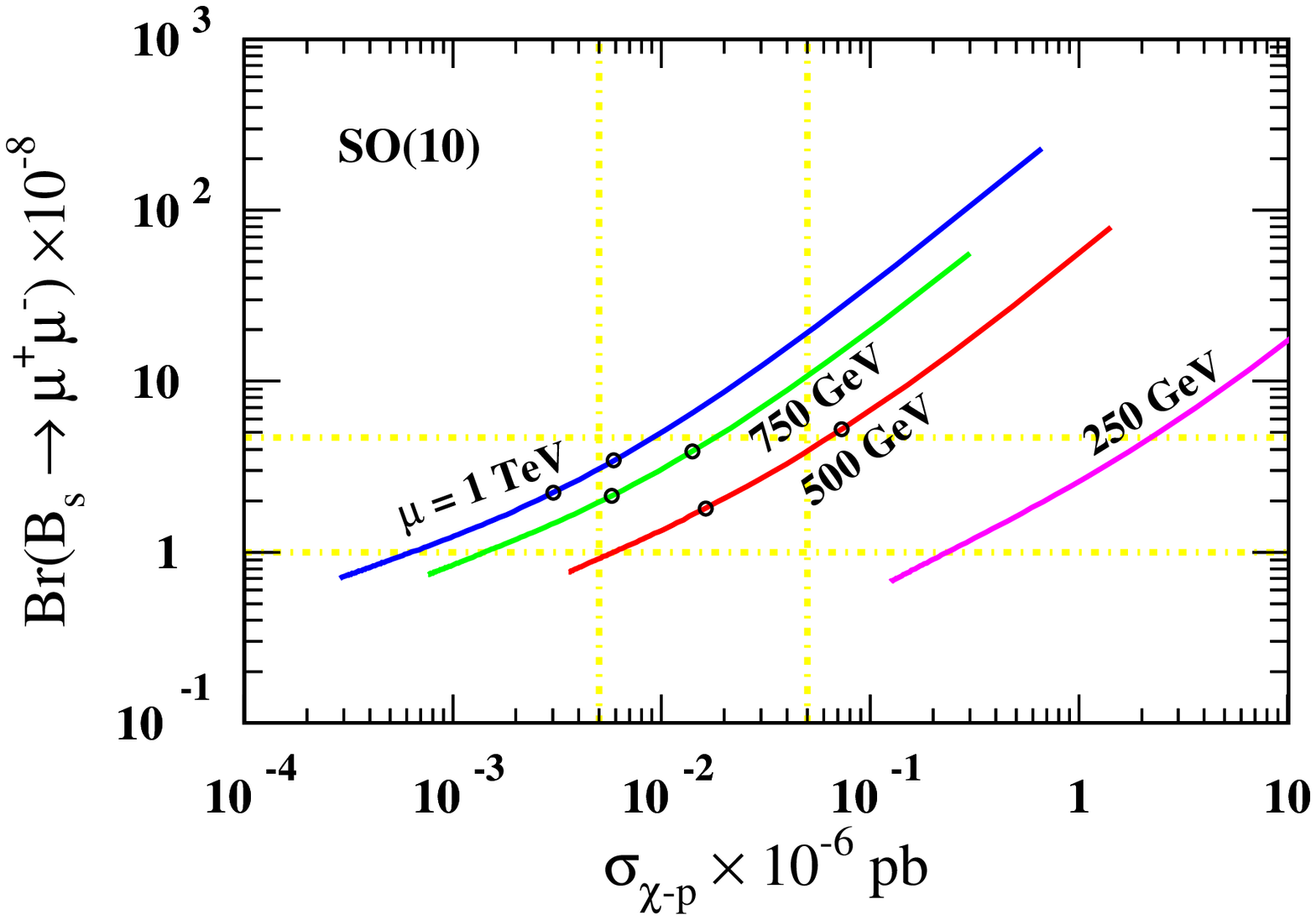}
 \caption{
Correlation between Br($B_s\to\mu\mu$) and $\sigma_{\chi - p}$ for the minimal-type
(a) SU(5) and (b) SO(10) cases, both for $\tan \beta = 40$, $A_0 = 0$, $m_{1/2} =
500$~GeV and $m_0 = 500$~GeV.
}
\end{figure}

\section{Conclusion}

We have investigated the GUT models
when the $B_s$-$\bar B_s$ mixing phase can become really large
as indicated in the Fermilab experiments.
We considered two cases:
one is the minimal-type of SU(5) model with type I seesaw.
The other is the minimal-type of SO(10) model with type II seesaw.
The difference between the two boundary conditions is whether there exists
a sizable off-diagonal element in the
left-handed squark mass matrix.
We emphasize that the sources of FCNC
in the GUT models  will be  restricted
if the large phase of $B_s$-$\bar B_s$ mixing persists in the upcoming result
from Fermilab.

In the case of a large $\tan\beta$,
the double penguin contribution dominates
the SUSY contribution to the $B_s$-$\bar B_s$ mixing
when the pseudo Higgs mass
is not too heavy.
Especially in the minimal-type of SU(5) model,
the pseudo Higgs mass should be low
enough to satisfy the experimental constraint
from $\tau\to\mu\gamma$ decay,
and because of this, the branching ratio of $B_s \to \mu\mu$ is sizable
and can be detected very soon.
The $\tau\to\mu\gamma$ constraint
also restricts the
Higgsino mass $\mu$ and the slepton masses,
and it may exclude some solutions
of the relic density of the neutralino dark matter.
In fact, the stau-neutralino coannihilation and the Higgsino dark matter solutions are not favored
for the large phase of $B_s$-$\bar B_s$ mixing in the
minimal-type of SU(5) model.
On the other hand,
the funnel solution
(in which the neutralinos annihilate through the heavy Higgs bosons pole)
is favored in this case.
For the funnel solution,
the pseudo Higgs mass is almost determined,
and the branching ratio of the  $B_s \to \mu\mu$ decay
is more predictive and can be observed soon for the values of soft masses which can be probed at the LHC.
The direct detection cross-section depends on the
heavy Higgs mass and the Higgsino mass, and correlated to
the $B_s \to \mu\mu$ in the case of
a large phase of the $B_s$-$\bar B_s$ mixing.
The dark matter-nuclear cross-section could be in the range to be detected very soon in the upcoming experiments in both SU(5) and SO(10) models.

In this paper, we have concentrated on the importance of
the 2nd and 3rd generation FCNC effects
such as Br($\tau\to \mu\gamma$) and $\phi_{B_s}$ correlation in GUT models,
since they can be correlated directly by the 23 mixing.
The constraints from Br($\mu\to e\gamma$) decay, $K$-$\bar K$ and $B_d$-$\bar B_d$
mixings 
 may also be important,
but these effects
depend on the details of
the flavor structure which can have a freedom of cancellation.
We refer to the Ref.\cite{Dutta:2006zt} for an analysis of
flavor violation including the first generation.

\section*{Acknowledgments}

This work 
is supported in part by the DOE grant
DE-FG02-95ER40917 (B.D. and Y.M.)
and the NSF grant No. PHY-0652363 (Y.M.).

\end{document}